# ssDNA sequencing by rectification

Ivana Djurišić[1], Miloš S. Dražić[1], Aleksandar Ž. Tomović[1], Marko Spasenović[2], Vladimir P. Jovanović[3] and Radomir Zikic[1,4]*

[1]Institute of Physics, University of Belgrade, Pregrevica 118, 11000 Belgrade, Serbia.
[2]Center of Microelectronic Technologies, Institute of Chemistry, Technology and Metallurgy, University of Belgrade, Njegoševa 12, 11000 Belgrade, Serbia.
[3]Institute for Multidisciplinary Research, University of Belgrade, Kneza Višeslava 1, 11000 Belgrade, Serbia.
[4]NanoCentre Serbia, Nemanjina 22-26, 11000 Belgrade, Serbia.

**Fast, reliable and inexpensive DNA sequencing is an important pursuit in healthcare, especially in personalized medicine with possible deep societal impacts. Despite significant progress of various nanopore-based sequencing configurations, challenges remain in resolution (due to thermal fluctuations or to sensitivity to molecular orientation) and speed, which are calling for new approaches. Here we propose a sequencing protocol for DNA translocation through a nanopore with side-embedded N-terminated carbon nanotube electrodes. Employing DFT and Non-Equilibrium Green's Function formalism, we show that the rectification ratio (response to square pulses of alternating bias) bears high nucleobase specificity. The rectification arises due to bias-dependent resistance asymmetry on the deoxyribonucleotide-electrode interfaces. The asymmetry induces molecular charging and HOMO pinning to the electrochemical potential of one of the electrodes, assisted by an in-gap electric field effect caused by dipoles at the terminated electrode ends. This sequencing protocol is sensitive, selective with orders of magnitude, has high resolution, and it is robust to molecular orientation.**

Next generation DNA sequencing is of great importance for applications ranging from personalized medicine through biotechnology to security issues[1]. Novel sequencing methods[2] need to be inexpensive, fast and reproducible, relying on real-time single-molecule readout[3]. With present biochemical approaches due to limits imposed on the read-out length, DNA is segmented into smaller pieces, which has an impact on the number of sequenced bases (length of reads), speed and cost of sequencing[2]. Apart from genomic sequencing, analysis of structural genetic variations in genomes (chromosomal translocations, inversions, large deletions, insertions, and duplications) necessitates reading the sequence of an entire chromosome from a single original DNA molecule[4]. Although significant research progress on long-reads[5] was demonstrated recently, a chromosome-sized sequence read length remains a challenge.

Nanopore-based approaches have emerged as a high-throughput DNA[6] and protein sequencing platform[3], where the variation of the ionic current during electrophoretic driving of single stranded DNA (ssDNA) decodes the sequence of the nucleotides along the entire length of a molecule[7,8,9]. Efficient DNA sequencing has been demonstrated only with biological nanopores[10,11,12], and even recently commercialized[5]. Solid-state nanopores offer some advantages over lipid-based pores, such as environmental stability and integration with CMOS technology. Typical thickness of solid-state nanopores is on the order of 10-20 nm, which is not thin enough for single-base resolution[13,14]. To increase resolution, researchers are integrating side-embedded[15] nano-electrodes (metals[16,17,18,19], carbon nanotubes (CNTs)[20,21,22], graphene ribbons[23,24,25,26,27,28,29], $MoS_2$[13,14,30]) with nanopores, to exploit the transverse tunnelling current through nucleotides, and to cross-correlate the two currents (ionic and transverse) for a more reliable read-out[14,19,25,27]. Currently, DNA sequencing by transverse current is technically very challenging, however with the rapid progress of nanofabrication, several experiments have already been reported[27,31,32,33].

A major drawback of transverse tunnelling current readout, especially in non-resonant transport when the electrochemical potential of the electrodes and the molecular levels are not matched, is the molecule misorientation with respect to the electrodes[17,26]. Imperfect alignment, occurring due to thermal fluctuations, causes nucleobase readout uncertainty. In resonant transport, molecular levels carry the current, increasing it by several orders of magnitude[20], and lessening the effect of molecule-electrode misorientation[26]. Resonant transport is often achieved by functionalization (termination) of electrodes with attachments ranging from complex[24,31] to single-atomic species such as nitrogen[20,23,34]. These species aim to involve HOMO and/or LUMO (highest occupied and lowest unoccupied molecular orbitals) in transport by shifting them towards the Fermi level through non-covalent interactions[20] such as hydrogen bonding[23] or $\pi$–$\pi$ stacking[31] with the termination species. Besides this approach, the energy of the molecular levels with respect to $E_F$ could be tuned with an electrostatic field generated by dipoles at electrode interfaces[35]. Other than tunnelling current, a measurement of rectification could also provide base pair fingerprints[36,37] although requiring sequencing of ssDNA.

Here we show at an *ab initio* level within DFT+NEGF at finite bias, that current rectification of ssDNA in N-terminated CNT nanogaps bears high nucleobase specificity and allows a new sequencing approach. The approach is sensitive and selective with orders of magnitude difference in amplitude between the different nucleobases. Also, since the rectifying ratio (the current ratio at negative and positive bias $RR=I(V)/I(-V)$) relies on HOMO-transport-channel on-off switching, the proposed approach is much more robust to molecular in-gap tilting than typical current amplitude measurements.

## Architecture for sequencing by rectification under alternating bias

The experimental configuration which we model is an electrically contacted single walled carbon nanotube (SWCNT) pair with a 15 Å gap, supported by a thin membrane (such as $Si_3N_4$) with a nanopore of a few nm in diameter. During ssDNA translocation through the nanopore, the rectifying ratio is measured via side-embedded SWCNT electrodes. We choose (3,3) SWCNTs as electrodes because of their small diameter comparable to the ~3.4 Å distance between two successive nucleobases of ssDNA (see Fig. 1, left panel). Deoxyribonucleotides (nucleotides) are single units of DNA, which besides the nucleobase (guanine, adenine, cytosine or thymine), comprise of deoxyribose sugar and one phosphate group. The DNA units are abbreviated as dGMP, dAMP, dCMP and dTMP (sketch of the structure shown in Fig. 1).

The central region for the calculations, also called the scattering region or the extended molecule, consists of a molecule (dGMP, dAMP, dCMP and dTMP), and a nitrogen (N)-terminated SWCNT at each side (Fig. 1, right panel). Electronic properties of electrodes outside the central region are not perturbed by the presence of the molecule. Transversal current through the nucleotides is calculated with DFT coupled with non-equilibrium Green's functions (NEGF), implemented in the TranSIESTA package[38]. In TranSIESTA, the Landauer-Büttiker equation yields electric current $I$ at a finite bias $V$ *as*:

$$I(V)=\frac{2e}{h}\int_{-\infty}^{+\infty}dE\left[f(E-\mu_L)-f(E-\mu_R)\right]T(E,V)\ , \qquad (1)$$

where $T(E,V)$ is the transmission spectrum, and $f(E-\mu_{L/R})$ is the Fermi-Dirac distribution in the left/right electrode with electrochemical potential $\mu_{L/R}=E_F\pm eV/2$ shifted from $E_F$ due to finite bias.

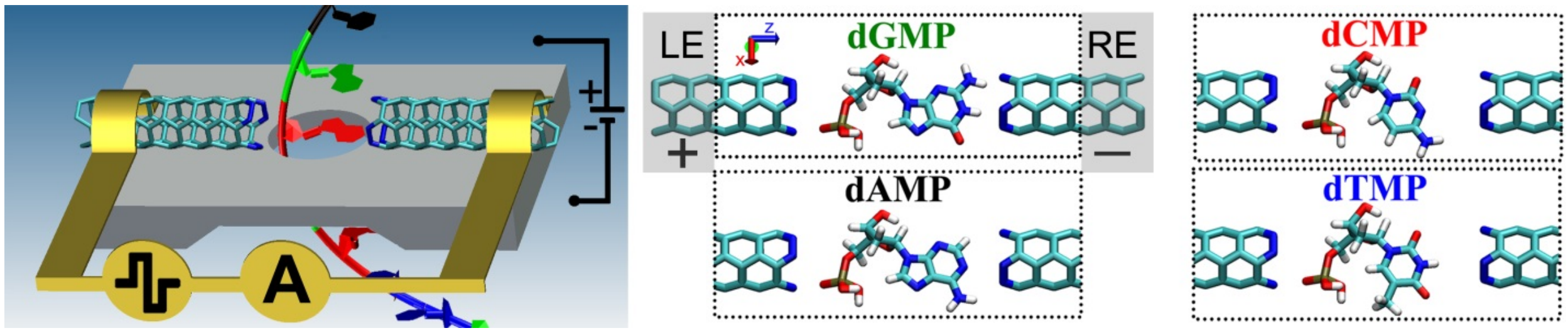

**Figure 1 | DNA sequencing set-up and system geometry.** The left panel shows the proposed set-up for DNA sequencing: ssDNA translocates through a nanopore with side-embedded N-terminated (3,3) carbon nanotube electrodes that serve to measure the rectifying ratio (transversal current) under square pulses of alternating bias. The right panel shows the geometry of the calculated system that consists of left and right electrodes (LE and RE) and a scattering region marked with a dotted rectangle which includes a nucleotide (dGMP, dAMP, dCMP and dTMP) and two SWCNT units with N termination. Arrows define Cartesian axes with the transport direction along the *z*-axis (blue). We adopt standard TranSIESTA notation: for negative bias the left electrode, facing a phosphosugar group, is positively charged (+ sign), i.e. LE has lower electrochemical potential than RE.

The detailed physical mechanism of the proposed sequencing is complex and we will discuss it in three steps: *i)* we describe the electrostatics of an empty nanogap with SWCNT electrodes; *ii)* we demonstrate rectification in electronic transport through nucleotides positioned between the electrodes; and *iii)* we show the discrimination between different nucleobases by the rectifying ratio.

## Empty nanogap: in-gap field effect engineered by polar bond termination

It is known that the electronic transport through a molecule in the nanogap could be increased by custom termination of the electrode ends. Here we consider the cases of N- and H-terminated SWCNT electrodes (Fig. 2).

Hirshfeld population analysis was employed in TranSIESTA[38] to explore the electrostatic properties of these configurations. It shows that negative charge accumulates in the N-termination layer, while positive charge goes to the adjacent layer of C atoms (Fig. 2a). As a result, a dipole moment pointing into the gap is created at the electrode interfaces. For H-termination, charge accumulation is opposite and the induced dipole moment points from the gap towards the electrodes (Fig. 2b). Qualitatively, the same charge distribution is obtained from Mulliken population analysis performed in NWChem[39], (see Fig. S1 in Supplementary Information (SI)). Charge distribution and dipole orientations remain almost unchanged in the presence of a nucleotide or under applied voltage, as depicted by dotted red lines in Fig. 2a,b for exemplar cases of dGMP at +1 V for N-termination and dGMP at 0 V for H-termination.

These dipoles of the polar bonds between the electrode and termination molecules generate a strong electrostatic field in the empty nanogap (Fig. 2c and Fig. S2 in SI). Depending on the bond dipole moment, the in-gap electrostatic field can be very strong (larger than 1.5 eV per electron for N-termination in Fig. 2c), causing the shift of molecular states with respect to the Fermi level, as in FET transistors (Fig. 2d). It is important to note that Hartree energy difference (see Fig. 2c) is in good agreement with calculated dGMP HOMO energy difference at zero bias between N- and H-termination (Fig. 2d).

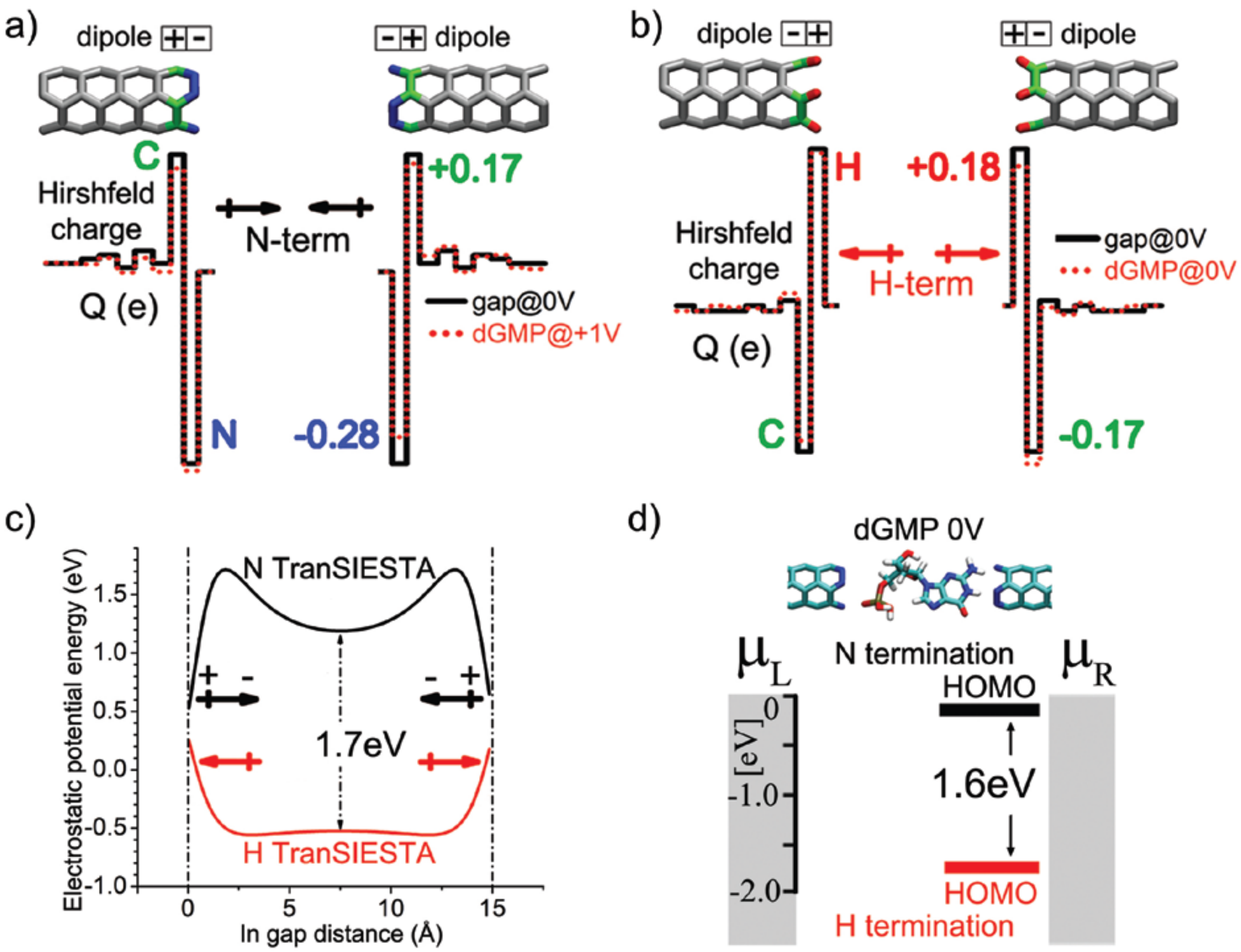

**Figure 2 | In-gap field effect induced by termination of CNT ends. a,** Hirshfeld charge excess *Q* at the ends of N-terminated CNTs without the nucleotide at 0 V (solid black line) and with dGMP at +1 V (dotted red line). The dipole arises in the N (blue) and the adjacent C atom layer (green). **b,** Hirshfeld charge excess *Q* at the ends of H-terminated CNTs without the nucleotide at 0 V (solid black line) and with dGMP at 0 V (dotted red line). An oppositely oriented dipole arises in the H (red) and the adjacent C atom layer (green). **c,** The zero-bias Hartree (electrostatic) potential energy obtained from TranSIESTA for N- (black) and H- (red) terminated nanogaps. **d,** Sketch of the energy difference between the HOMO level of dGMP and $E_F$ calculated in the scattering region depicted on top at zero bias for N- and H-termination (black and red bars, respectively).

## Deoxyribonucleotides in the nanogap: charge distribution and HOMO pinning

The alignment of the HOMO level with the electrochemical potential (due to the in-gap electric field generated by the CNT termination) is important for the current flow through the single molecule. It is necessary to emphasize that due to the confined geometry (small distances) the molecule and the electrodes cannot be treated independently. Their coupling induces broadening of the molecular level, and the overlap of their wave-functions will cause charge redistribution on the molecule, with serious consequences for the transmission between the electrodes. For a fixed geometry of the nanopore and CNT electrodes, the coupling is nucleobase dependent, since the nucleobases have different sizes. For example, the dGMP HOMO level, with the help of the in-gap electric field gets to the vicinity of the Fermi level ($E_F \equiv \mu_{R,L}$ at $V=0$), $\Delta = E_F - E_{\mathrm{HOMO}} = 60$ meV, (see Fig. 3a, top panel) and the transport through the molecular state dominates over coupling-induced broadening. However, for positive polarity, the dGMP HOMO follows the electrochemical potential $\mu_R$, i.e. keeps the energy distance to $\mu_R$ almost constant ($\Delta$) independently of the applied bias (see Fig. 3a), hence the molecular level will not contribute to transport as would be expected from the widely used zero-bias limit transport theory $T(E,V) \approx T(E,0)$. This effect, illustrated in Fig. 3a middle and lower panels, is shown in Fig. 3b for a range of applied biases. Consequently, HOMO is swept through the "integration window" (marked yellow in Fig. 3a), with applied bias. This bias-polarity-triggered on-off switching of the HOMO-transport-channel results in strong rectification (*RR*=50@100mV).

To explore the physics of the HOMO pinning to $\mu_R$, we employ again Hirshfeld's population analysis, calculating bias dependence of the charge excess $Q$ on a nucleotide. We find that finite charge excess resides on the nucleotide for all bias levels (see Fig. 3c). This charge appears as a result of charge redistribution between the molecule and electrodes, while the system remains electroneutral. Note that the considered system is not in the weak coupling (quantum dot) regime and that the electronic wave function is spatially distributed both on the nucleotide and electrode atoms, resulting in partial charge (fraction of elementary charge) residing at the nucleotide. The molecular charging energy is[40,41]:

$$E_C = -eq/C_{ES}, \qquad (2)$$

where $e$ is the elementary charge, $q=Q(V)-Q(V=0)$, $Q(V)$ is the charge excess at the nucleotide and $C_{ES}$ is the electrostatic capacitance across the system. In our model the charging energy is correlated with the shift of the $E^{\mathrm{HOMO}}$ with field in respect to the zero-bias position. Fitting $E_C = E_{\mathrm{HOMO}}(V) - E_{\mathrm{HOMO}}(V=0)$ gives $C_{ES} = 0.13$ e/V (Fig. 3d). This implies that charging shifts the energy level with respect to $E_F$[42,43]. Even weak charging of ~0.1 e for dGMP at 1.6 V shifts HOMO by 800 meV (see Fig. 3c and d). We emphasize that HOMO shifting with $E_F$ by over 1 eV cannot be due to hydrogen bonding that has energy limited to about 300 meV, instead the dominant mechanism is charging of the nucleotide. Consequently, the lowering of the energy level of the molecule with respect to $E_F$ keeps the energy difference to $\mu_R$ constant and bias independent, mimicking a "pinning" to the electrochemical potential (Fig. 3b).

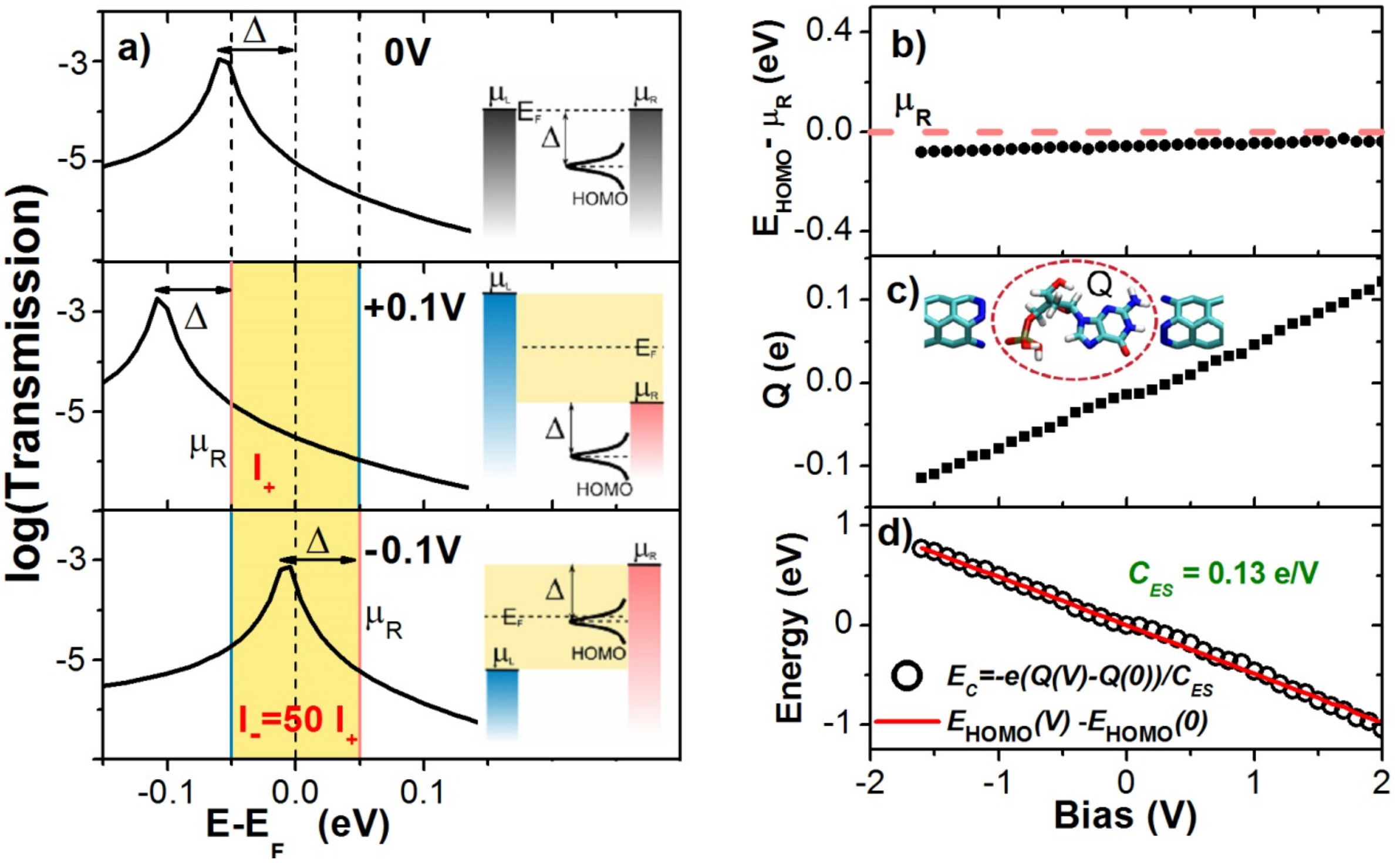

**Figure 3 | Charge transport through dGMP, HOMO pinning, charge excess and charging energy. a,** Semi-log plots of electronic transmission with respect to Fermi energy at different bias. **b,** Energy difference between HOMO and electrochemical potential of the right electrode $\mu_R$ in respect to bias. **c,** Charge excess $Q$ obtained from Hirshfeld analysis versus bias. **d,** difference of HOMO energy level at finite and zero bias (circles), and charging energy $E_C$ obtained from equation (2) for $C_{ES}$ = 0.13 e/V (solid line). All calculations are performed in TranSIESTA.

To explore the origin of molecular charging, we calculate the electrostatic potential energy profile across the gap with a molecule. We will show the relation between *charging-induced HOMO pinning and bias-dependent changes in potential energy profile* on the example of dAMP. The situation is more subtle for smaller nucleotides (dAMP, dCTP, dTMP), where the HOMO level does not pin at all biases to the electrochemical potential (see Fig. 4a and Figs. S3-7 in SI). We call the "weak pinning" regime the region in which HOMO shifts independently of $\mu_R$ with bias voltage. At a threshold voltage where HOMO shifts to near $\mu_R$ the hybridization between the molecule and electrode wave functions launches the charging of the molecule. In this range $E_{HOMO}$-$\mu_R$ becomes bias independent (Fig. 4a) bringing the system into the "strong pinning" regime.

In weak pinning when HOMO does not follow the electrochemical potential, charging of the nucleotides is almost negligible (Fig 4b) and it sets in only at higher biases. This is valid for all nucleotides (see Fig. 4b and Fig. S7 in SI). Again, the charging energy is correlated with the energy difference of HOMO in weak and strong pinning regimes (Fig. 4c and Fig. S8 in SI). Using an extension of equation (2), $q$ is defined as $q=Q(V)-Q^{Weak}(V)$, where $Q^{Weak}$ and $E_{HOMO}^{Weak}$ are linear extrapolations of $Q$ and $E_{HOMO}$ from the weak pinning regime, respectively (solid purple lines in Fig. 4b,a). The fit of $E_C=E'_{HOMO}$ (shown as a red line in Fig. 4c), where $E'_{HOMO} = E_{HOMO}(V) - E_{HOMO}^{Weak}(V)$ yields $C_{ES}$ = 0.25 e/V.

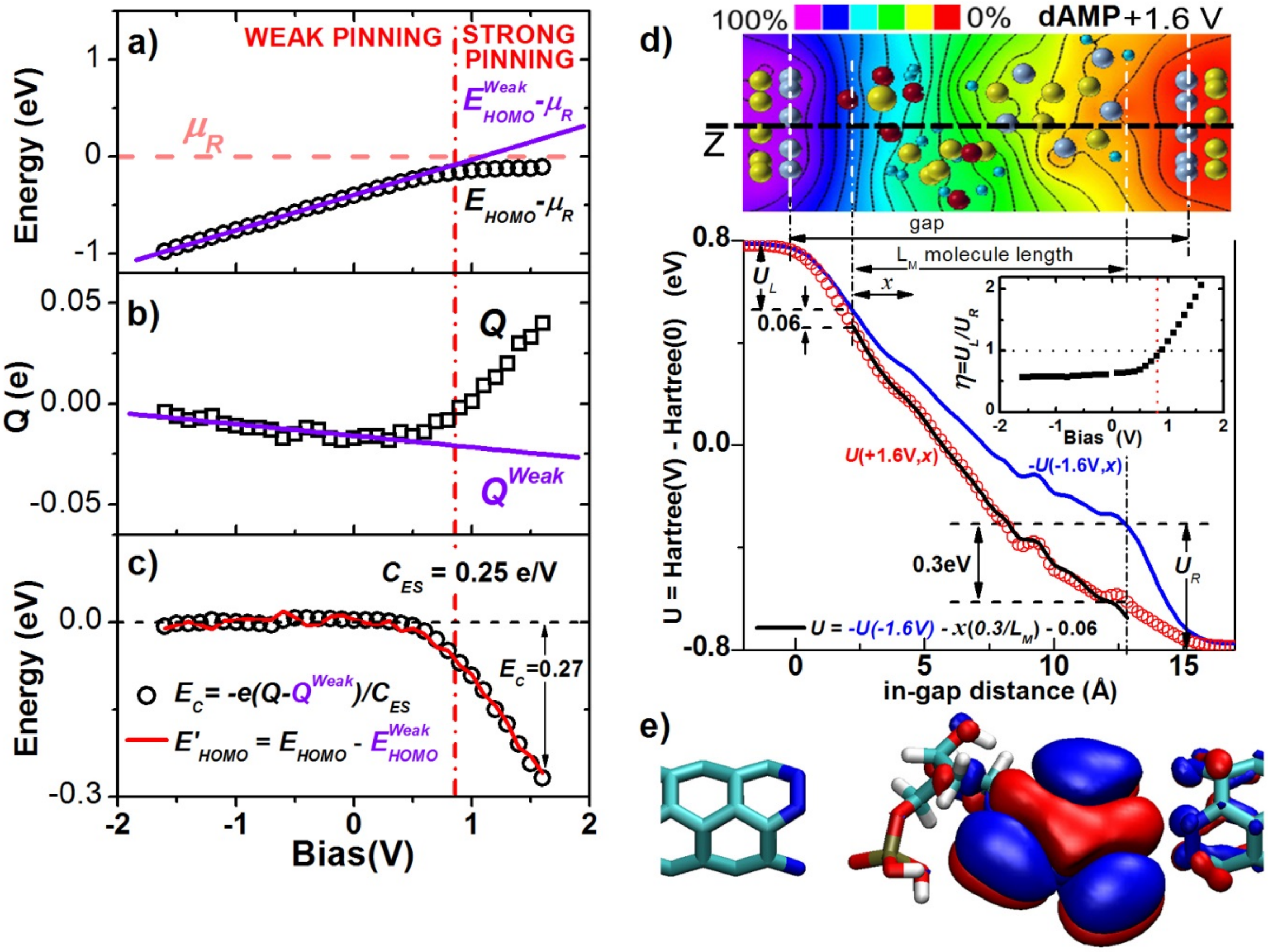

**Figure 4 | Transport through dAMP, voltage dependent charge excess and potential drop asymmetry at interfaces. a,** Open black circles - energy difference between HOMO and electrochemical potential of right electrode $\mu_R$. Solid purple line – linear extrapolation of the weak pinning regime $E_{HOMO}^{Weak}$. **b,** Charge excess *Q* obtained from Hirshfeld analysis versus bias (open black squares). Solid purple line – linear extrapolation of the weak pinning regime ($Q^{Weak}$). **c,** $E_{HOMO} - E_{HOMO}^{Weak}$ (circles), charging energy $E_C$ from extension of equation (2) for $C_{ES}$ = 0.25 e/V (solid red line). **d, Top:** Hartree electrostatic potential energy difference *U(V)-U(0)* normalized to [0,100%] at +1.6 V in the plane of the nucleobase (nitrogen atoms are grey, carbon atoms are yellow) showing equipotential lines. **Bottom:** *U(V)-U(0)* extracted along the *z*-direction (dashed black line in top panel), for -1.6 V (solid blue line) and +1.6 V (open red circles). For clarity the blue curve at negative bias is -(*U*(V)-*U*(0)). $U_L$ and $U_R$ are potential drops at left and right interfaces. $L_M$ =10.62 Å is the length of the molecule. The difference $U_C$ between blue and red curves at the right interface is found to be 0.3 eV. The solid black line $U = -U(-1.6) - xU_C/L_M - 0.06$ is a linear transformation of *-U(-1.6)* (blue curve). **Inset:** asymmetry factor $\eta = U_L / U_R$ as a function of bias in weak and strong pinning regimes. **e,** Spatial distribution of the wavefunction of dAMP HOMO level at zero bias.

We now calculate the potential energy in the plane that contains the nucleobase (Fig. 4d). The molecule distorts the shape of equipotential lines, which would otherwise follow a smooth profile (top panel Fig. 4d). The bottom panel of Fig. 4d depicts the Hartree potential energy *U(x)* at -1.6 V (blue solid line – weak pinning) and +1.6 V (red circles – strong pinning) voltages offset by Hartree potential at zero bias. For clarity, for negative bias, –*U*(-1.6) is shown instead of *U*(-1.6). Here we find the difference between potential drops at the right interface in the strong and weak pinning regimes $U_C$ to be equal to 0.3eV. The potential profile across the molecule in the strong pinning regime can be reconstructed from the weak pinning profile by adding a term that corresponds to the classical potential drop in an infinite parallel plate capacitor ($V=\mathcal{E}d$):

$$U^{Strong} = U^{Weak} - \frac{U_C}{L_M}x - \Delta U_L \qquad (3)$$

where $\Delta U_L$ is an offset (small potential drop difference) at the left interface, $L_M$ is the total length of the molecule along the $z$-direction, and $\mathcal{E}=U_C/e/L_M$ is electric field strength, i.e. $U_C$ is energy lost per electron along the entire molecule. In other words, during tunnelling from one to another electrode in the strong pinning regime, the electron effectively experiences potential energy loss ($\Delta U = e\mathcal{E}x$) along a constant parallel electric field $\mathcal{E}$. In the example of dAMP at 1.6 V, shown in Fig. 4d, the energy loss $U_C$ =0.3 eV (energy difference between weak and strong pinning at the right interface) is in good agreement with both $E'_{HOMO}$ and the molecule charging energy ($E_C$) that were found to be 0.27 eV and 0.26 eV, respectively (Fig. 4c), while $\Delta U_L$=0.06 is significantly lower than these energies. Furthermore, the same rule for in-molecule potential energy profile reconstruction from weak (not charged) to strong (charged) pinning by adding capacitor-like behaviour (with energy loss per electron $U_C \approx E'_{HOMO}$) is found to be applicable for all biases (for more details see Fig. S9 in SI).

One should notice that there is an asymmetry in the potential drop on the left and right electrodes (marked as $U_L$ and $U_R$, respectively) as shown in the inset to Fig. 4d by introducing $\eta = U_L / U_R$ as the asymmetry factor. The vertical blue dashed line indicates the transition to strong pinning, as in Fig. 4b. We observe that for weak pinning $\eta$ is bias-independent, in contrast to the strong pinning regime.

To explore the mechanism that causes asymmetry in the potential drop at the interfaces we focus on the spatial distribution of the HOMO orbital. Its wave function for the dAMP at zero bias is shown in Fig. 4e (for other nucleotides it is given in Fig. S9 of SI). There is significant spatial extension of the HOMO wave function and an overlap with the atomic orbitals of the right electrode. For bias values at which, in addition to the spatial overlap, there is a matching between dAMP HOMO energy and the electrochemical potential of the right electrode, the contact resistance decreases, resulting in a considerable lowering of the potential drop at the right interface (Fig. 4d). Thus, whenever there is a strong asymmetry in the potential drop at the interfaces, the system compensates the ensuing energy difference $U_C$ by lowering molecular energy via charging of the molecule. This asymmetry appears due to overlap, hybridization[44] of the HOMO and Bloch states from one of the electrodes in energy and real space.

## Nucleobase specificity of rectification

Due to the strong asymmetry between weak and strong pinning regimes, there is an important difference in current flow at two opposite bias values, in other words there is rectification. In our notation (Fig. 1), for dGMP negligible current flows when the electrode facing the phosphosugar group is at a lower potential than its counterpart, while a much larger current flows under opposite bias. This rectification occurs even at small bias. For example, at 100mV, the current is fifty times larger for negative than for positive bias (Fig. 3a).

This strong rectification is caused by two mechanisms. First, transport through the dGMP HOMO molecular state dominates total transport, and second, the HOMO level is “pinned” to the electrochemical potential by molecular charging. Since charging occurs through hybridization of the HOMO level and the electrode Bloch states it is sensitive to the size of the nucleobase molecules.

In consequence, this strong bias-dependent asymmetry, which is nucleobase-dependent, can be exploited for sequencing. Here we propose a sequencing protocol where constant negative bias is applied to the electrodes to keep the phosphosugar group facing the left electrode, as sketched in the right panel of Fig. 1. The protocol is performed by interrupting the constant bias with short square pulses of +0.8 V to obtain an optimal *RR*, as depicted in Fig. 5a. The calculated rectifying ratio from the pulse period (PP) for all four nucleotides as a function of the absolute value of the constant bias is shown in Fig. 5b. It is evident that the nucleotides exhibit different rectification behaviours, which makes them distinct for sequencing. The most extreme differences in *RR* between the different nucleotides are at a bias of 0.8 V, designated as the optimal working point for the sequencing protocol. At this bias, *RR* is 1,000, 50, 1 and 0.3 for dAMP, dGMP, dCMP and dTMP, respectively.

While the transverse current amplitude is highly sensitive to mutual molecule-electrode orientation[17,26] (Fig. 5c and Fig. S10 in SI), rectification strongly depends on nucleotide HOMO position with respect to $E_F$ and is less sensitive to orientation. Calculations for dGMP oriented at an angle of 30° with respect to the y-axis (tilted in the plane which contains the nucleotide base) at ±0.2 V bias show that while transmission is diminished ($I_{\mathrm{rot0}}/I_{\mathrm{rot30}}$ = 100) due to tilting, the position of HOMO with respect to $E_F$ remains the same (see Fig. 5c) with *RR* changing only slightly. Our results suggest that the rectifying ratio is robust to the effect of orientation ($RR_{\mathrm{rot30}}$=6$RR_{\mathrm{rot0}}$) and we propose measuring *RR* of single-stranded ssDNA in a high-throughput sequencing tool. In the present approach we do not consider environmental effects (water molecules and counter ions) on electronic transport properties of nucleotides. The presence of water molecules and counter ions would on average exert a similar effect on the different nucleotides, thus our estimate is that environmental effects would not perturb our protocol[45].

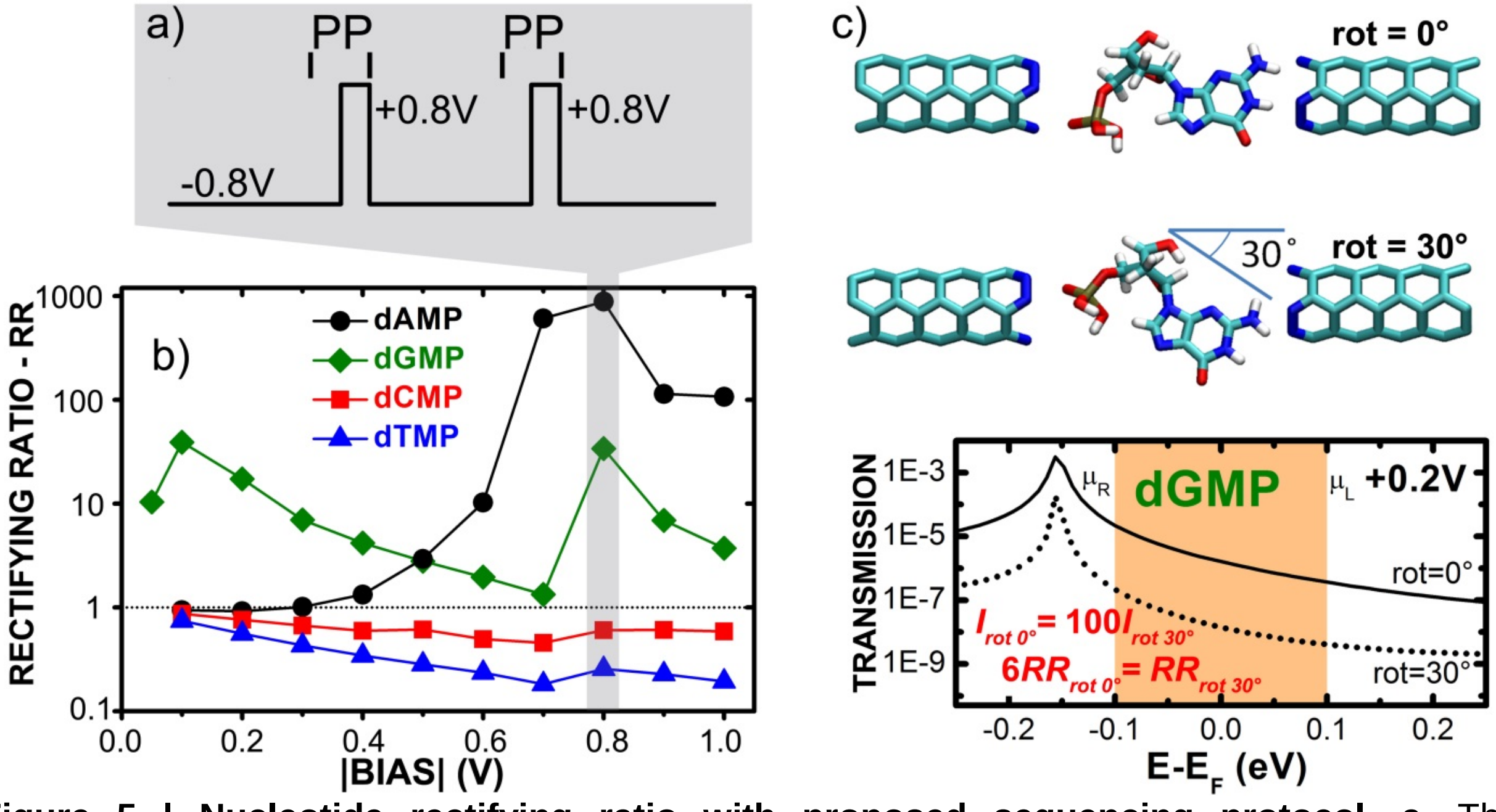

**Figure 5 | Nucleotide rectifying ratio with proposed sequencing protocol. a,** The suggested protocol to distinguish between nucleotides by rectifying ratio (*RR*): interrupt a constant bias of –0.8 V with short square pulses of +0.8 V and during this alternating square pulse period ("PP") measure the rectifying ratio. The rectifying ratio is defined as the absolute value of the ratio of currents for negative and positive bias. **b,** Calculated *RR* for different nucleotides (dGMP – diamonds, dAMP – circles, dCMP – squares and dTMP – triangles) as a

function of absolute value of applied bias. **c,** Sketch of molecule-electrode tilt for $0^\circ$ and $30^\circ$. **d,** Transmission of dGMP as a function of energy at +0.2 V for two different molecule-electrode tilt angles.

## Conclusions and outlook

We have shown here, based on DFT+NEGF at finite bias approach, strong rectification of transverse current through a probed ssDNA nucleotide (molecule) in a (3,3) single wall carbon nanotube nanogap, which could be a suitable quantity for single molecule read-out. The rectification is achieved by on-off switching of electronic transport through the HOMO level, sustained by partial charging of the molecule, which is generated by asymmetric hybridisation of the HOMO with Bloch states from one of the electrodes.

We have also identified that functionalising electrode ends customizes a dipole-induced strong perpendicular in-gap electric field (~1eV/nm) that can shift the molecular levels (HOMO or LUMO) towards Fermi level, generating high transmission and rectification along with the hybridisation. Although dipoles from electrode ends strongly rely on electrode-species polar covalent bonding, our approach offers a platform for tailoring the in-gap local gating by customizing electrode's termination. For example in semiconductor-based nano-junctions for energy harvesting and optoelectronic, and in graphene-nanopore based high-spatial-resolution system for single-molecule protein sequencing and probing protein secondary structure.

## Methods

Transversal current through the nucleotides is calculated with density functional theory (DFT) coupled with non-equilibrium Green's functions (NEGF), implemented in the TranSIESTA package[38]. The basis set was double zeta polarized for all atoms, while the exchange-correlation functional was approximated with the Perdew-Burke-Ernzerhof (PBE) functional[46]. Core electrons were described with Troullier-Martins norm-conserving pseudopotentials[47]. For electrode calculations we used 1×1×64 k-points while in the scattering region only the Γ-point was considered. The mesh cutoff value was 170 Ry for the real-space grid. Geometry of the electrode's unit cell is cylindrical with 2.041Å radius and 2.468Å lattice constant[48]. Geometry of N-termination, as well as the geometry of four nucleotides deoxyguanosine monophosphate (dGMP), deoxyadenosine monophosphate (dAMP), deoxycytidine monophosphate (dCMP), deoxythymidine monophosphate (dTMP) was relaxed using SIESTA[49].

## Acknowledgments

This work was supported by the Serbian Ministry of Education, Science and Technological Development through projects 171033 and 41028. We gratefully acknowledge financial support from the Swiss National Science Foundation (SCOPES project No.52406) and the FP7-NMP, project acronym *nanoDNAsequencing*, GA214840.

SUPPLEMENTARY INFORMATION

# ssDNA sequencing by rectification

Ivana Djurišić[1], Miloš S. Dražić[1], Aleksandar Ž. Tomović[1], Marko Spasenović[2], Vladimir P. Jovanović[3] and Radomir Zikic[1,4]*

[1]Institute of Physics, University of Belgrade, Pregrevica 118, 11000 Belgrade, Serbia.
[2]Center of Microelectronic Technologies, Institute of Chemistry, Technology and Metallurgy, University of Belgrade, Njegoševa 12, 11000 Belgrade, Serbia
[3]Institute for Multidisciplinary Research, University of Belgrade, Kneza Višeslava 1, 11000 Belgrade, Serbia.
[4]NanoCentre Serbia, Nemanjina 22-26, 11000 Belgrade, Serbia

radomir.zikic@ncs.rs
∗To whom correspondence should be addressed

## S1. Dipole formation by N-termination

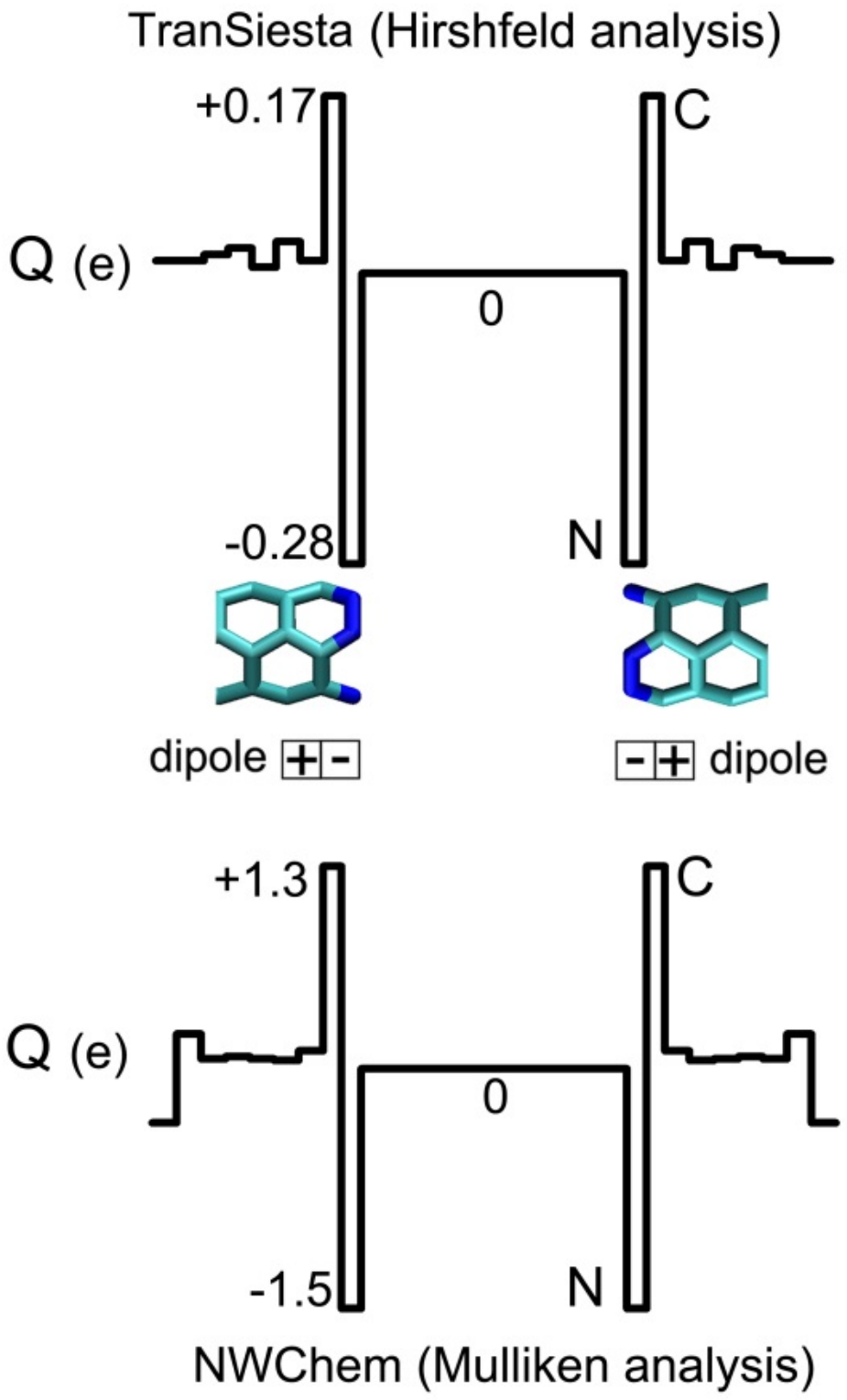

**Figure S1. Charge excess *Q* of different atomic layers of N-terminated (3,3) carbon nanotubes** calculated with TranSIESTA (top panel) and NWChem (bottom panel) using Hirshfeld and Mulliken population analyses, respectively, showing that nitrogen passivation creates dipoles at the electrode ends.

Charge distribution for N-terminated (3,3) CNTs was calculated with the TranSIESTA[1]and NWChem[2] packages. The results from the two packages are in good qualitative agreement, as they both predict formation of dipoles at the N-terminated ends of CNTs.

**S2. Modelling of in-gap electrostatic potential energy**

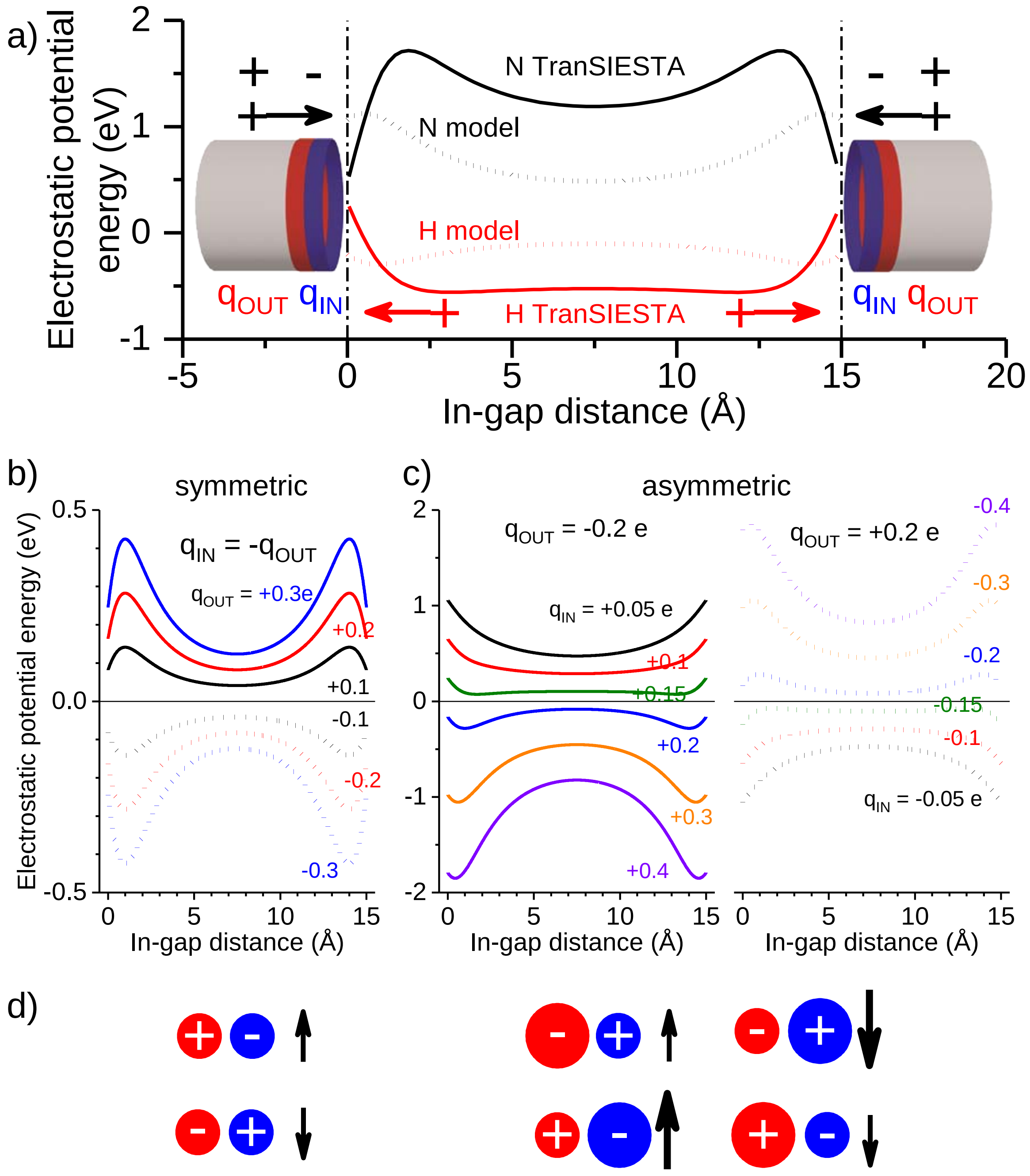

**Figure S2. a) Modelled electrostatic potential energy of oppositely oriented dipoles** made of two rings (blue and red) each. In the N (H) model the blue ring carries homogeneous linear charge $q_{IN}$ of -0.28 e (+0.18 e) and the red ring carries $q_{OUT}$ of +0.17 e (-0.17 e). The diameter of the rings is 4 Å. The distances between red and blue rings are 1.124 Å and 0.92 Å for the N and H models, respectively. Arrows indicate the orientation of dipoles in the N (black) and H (red) model. The zero-bias Hartree potential energy $U_H(0)$ obtained from TranSIESTA [1] for N- (black solid line) and H- (red solid line) terminated nanogaps is also given for comparison. b) Electrostatic potential energy from the symmetric dipole model, i.e. $q_{IN}$ = - $q_{OUT}$, taking that $q_{OUT}$ carries charge from -0.3 e to +0.3 e. c) Electrostatic potential energy from the asymmetric dipole model i.e. $q_{IN} \neq - q_{OUT}$. In the left panel $q_{OUT}$ is negative (-0.2 e) and $q_{IN}$ varies from +0.005 e to +0.4 e, while in the right panel $q_{OUT}$ is positive (+0.2 e) and $q_{IN}$ varies from -0.05 e to -0.4 e. In b) and c) the diameter of the rings is 4 Å and their mutual distance is 1 Å. d) In the symmetric case ($q_{IN}$ = - $q_{OUT}$, left panel), positive $q_{OUT}$ (red circles) raises in-gap potential energy (upward arrow), while $q_{IN}$ positive decreases the energy (downward arrow). In the asymmetric case (right panel), the sign of the larger charge determines rising (-) or lowering (+) of the energy. The effect is more pronounced (larger arrows) for larger $q_{IN}$ (blue circles).

A simple model was used to calculate electrostatic potential energy in the gap along the cylindrical symmetry axis: the layer of N and adjacent layer of C atoms were represented with charged rings, with separation and diameter defined by the actual CNT geometry (Fig. S2a). Rings bare homogeneous linear charge which is obtained from Hirshfeld population analysis from TranSIESTA[1] for the cases of N- and H-terminated gaps. Two opposite dipoles at CNT ends create a saddle-shaped electrostatic potential energy profile (dotted black and red lines in Fig. S1a) with spatial variations as large as 1 eV, indicating that a molecule placed in the nanogap will experience a strong field effect. Finally, the in-gap Hartree potential obtained from TranSIESTA for N- and H- (Fig. S2a black and red solid lines respectively) terminated nanogaps is in good qualitative agreement with the results obtained from our simple model, indicating that the origin of the potential landscape is related to dipoles. The dipoles arise from strong C-N polar bonds, providing a platform for tailoring the in-gap field effect by customizing termination species.

The model shows that for a "symmetric" dipole ($q_{IN}$ = - $q_{OUT}$) the field effect is stronger for a larger dipole moment (Fig. S2b). In this case, when the outer red ring is positively charged (solid lines in Fig. S2b) electrostatic energy is positive, indicating that a molecule placed in the nanogap will experience a positive shift of energy levels (Fig. S2d). The opposite is true for a negatively charged outer ring (dotted lines in Fig. S2b).

The "asymmetric" dipole case ($q_{IN} \neq$ - $q_{OUT}$), which more correctly represents real N-terminated CNTs, has more complicated behaviour. Whether the electrostatic energy will be positive or negative mainly depends on absolute values of $q_{IN}$ and $q_{OUT}$: irrespective of orientation of the interface dipole, **the electrostatic energy in the gap rises whenever negative charge dominates over positive and vice versa** (Fig. S2c,d). The larger the difference between absolute values of $q_{IN}$ and $q_{OUT}$ (asymmetry) the larger the shift of electrostatic potential (Fig. S2d). The gap size, the distance between dipole-forming rings and the diameter of the rings can also influence the behaviour of the energy, but to lesser extent.

## S3. Participation of molecular levels in transmission

Electronic transmission through nucleotides for different bias levels is depicted in Figs. S3-S6 with focus on transmission peaks that significantly contribute to the current (dots), i.e. extended molecule levels that determine the shape of the *I-V* curve. The main contribution to current comes from the nucleotides' HOMO (red dots), while in the case of dAMP and dGMP some extended molecule levels localized on the left electrode (blue and green dots) also participate in transport. Whenever a transmission peak with significant intensity enters the bias window (light blue and orange shaded regions in Figs. S3-S6) there is an increase in current. Note that the HOMO of dGMP strictly follows the electrochemical potential of the right electrode – HOMO is strongly pinned at all bias values. This is also true for the HOMO of dAMP (dCMP) for bias between +0.8 and +1.6V (-0.9 and -1.6V). Strong pinning of dTMP HOMO level starts at -1.6V. For other bias values only weak pinning of HOMO is observed (for example, for dCMP at positive bias in the right panels of Fig. S4, the red dot barely changes position with bias).

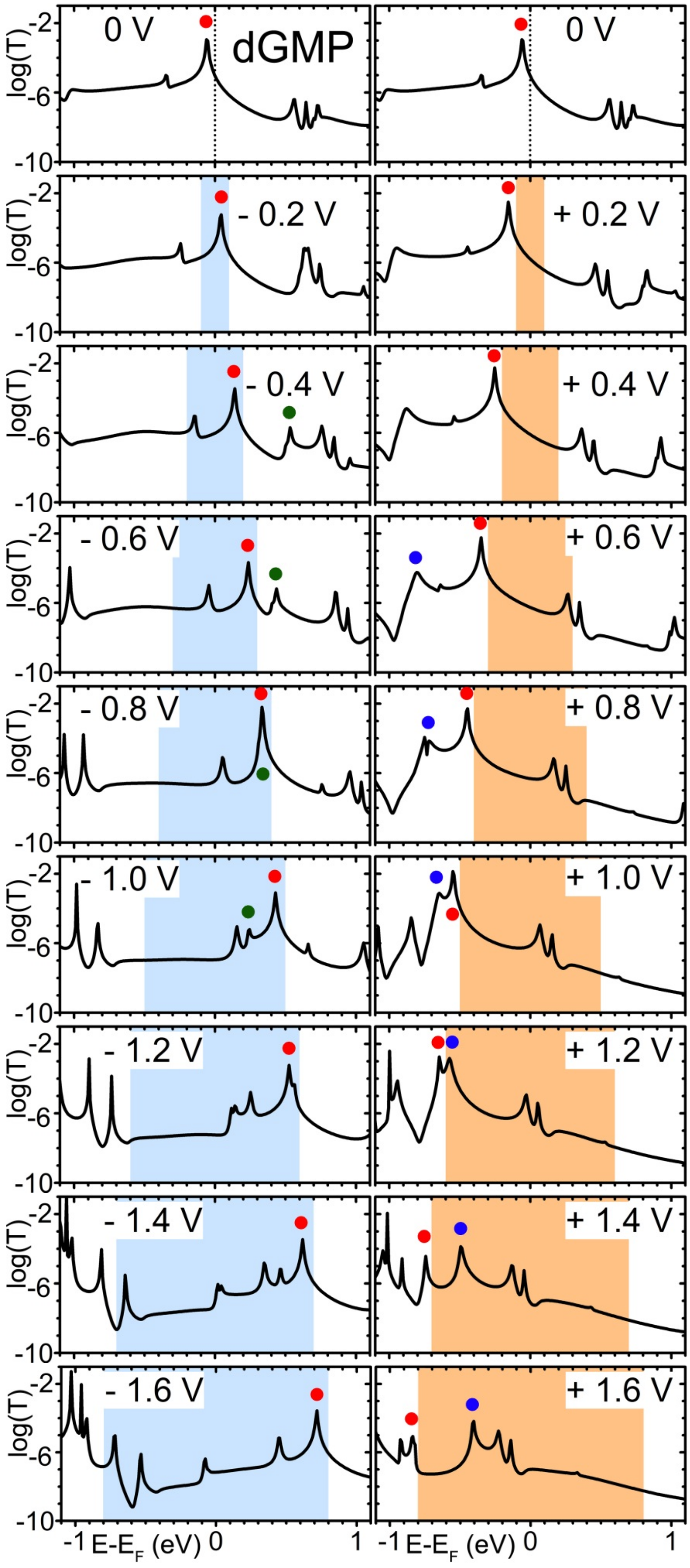

**Figure S3. Electronic transmission vs. $E$–$E_F$ for different bias values of dGMP between two N-terminated (3,3) carbon nanotubes**. In left (right) panels bias window is shaded light blue (orange) for negative (positive) bias. Red spot marks the position of the dGMP HOMO transmission peak, while blue and green spots mark the transmission peaks corresponding to energy levels of the extended molecule localized on the left electrode.

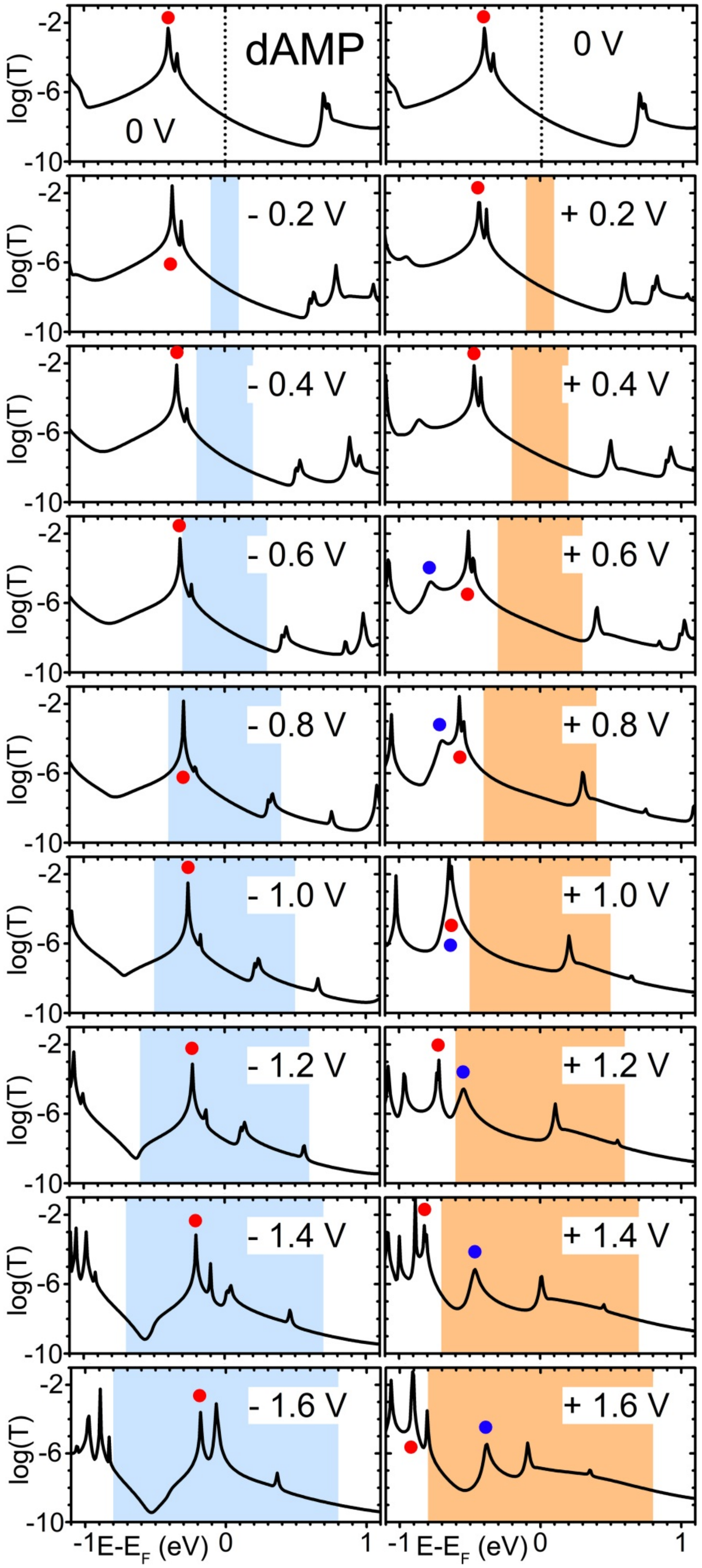

**Figure S4. Electronic transmission vs. $E$–$E_F$ for different bias values of dAMP between two N-terminated (3,3) carbon nanotubes**. In left (right) panels bias window is shaded light blue (orange) for negative (positive) bias. The red spot marks the position of the dAMP HOMO transmission peak, while the blue spot marks the transmission peak corresponding to an energy level of the extended molecule localized on the left electrode.

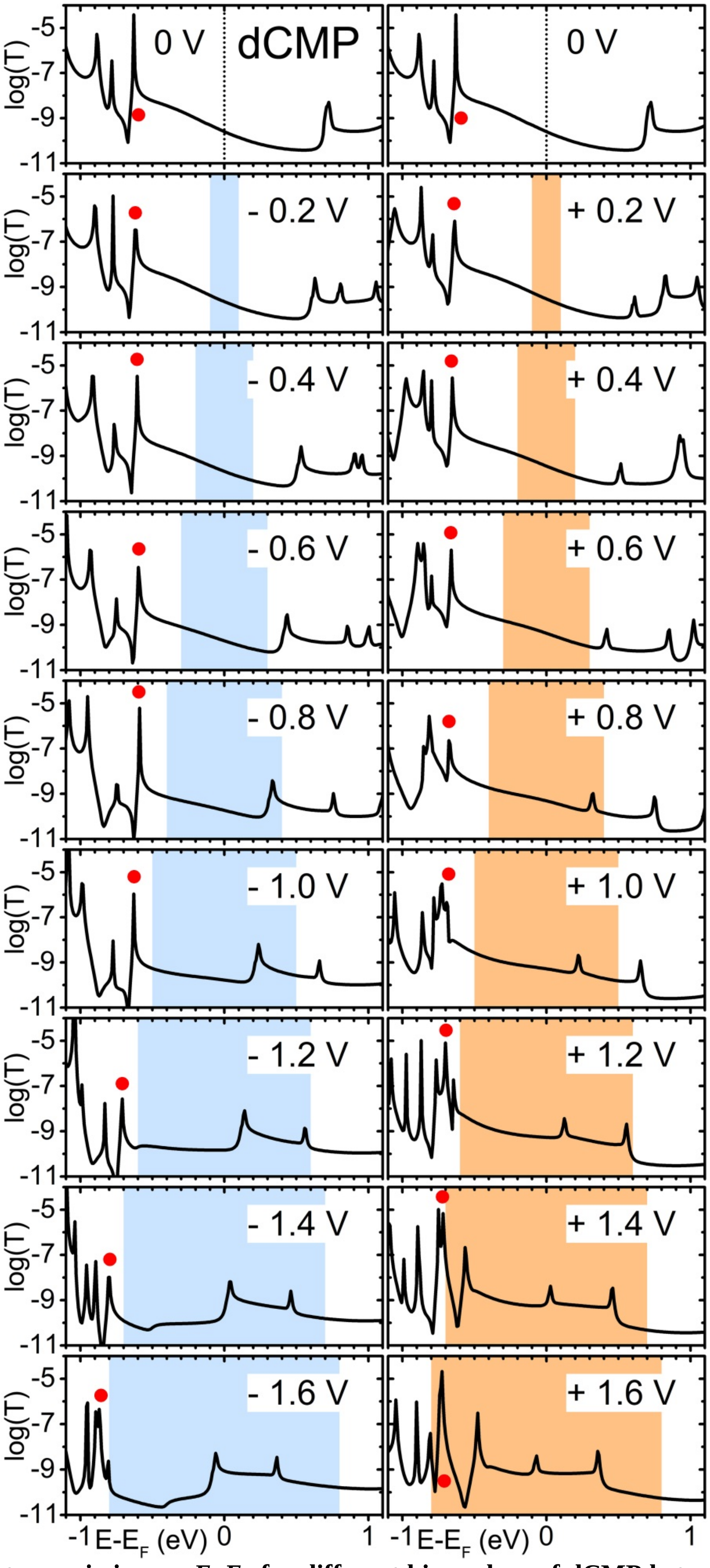

**Figure S5. Electronic transmission vs. *E–E$_F$* for different bias values of dCMP between two N-terminated (3,3) carbon nanotubes**. In left (right) panels the bias window is shaded light blue (orange) for negative (positive) bias. The red spot marks the position of thedCMP HOMO transmission peak.

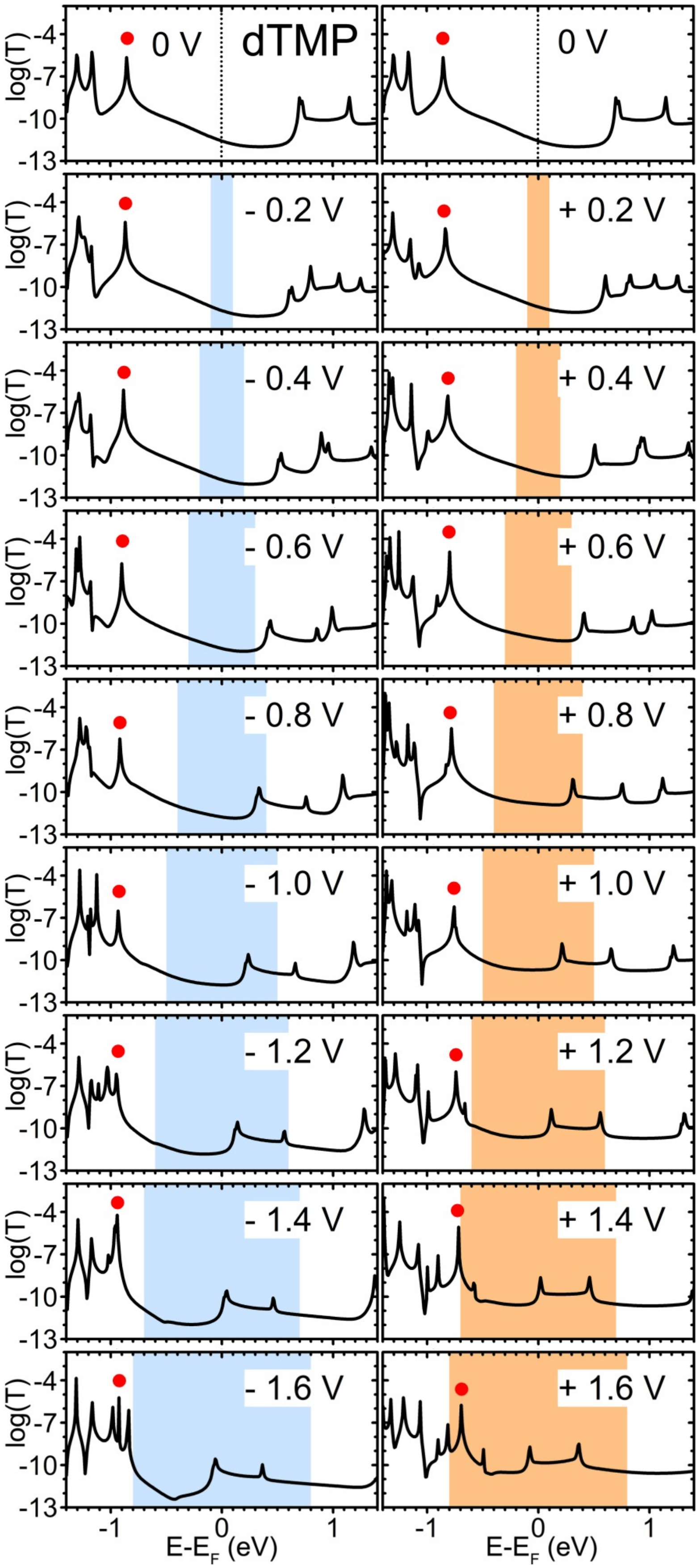

**Figure S6. Electronic transmission vs. *E–E*$_F$ for different bias values of dTMP between two N-terminated (3,3) carbon nanotubes**. In left (right) panels the bias window is shaded light blue (orange) for negative (positive) bias. The red spot marks the position of the dTMP HOMO transmission peak.

## S4. HOMO pinning and molecular charge excess

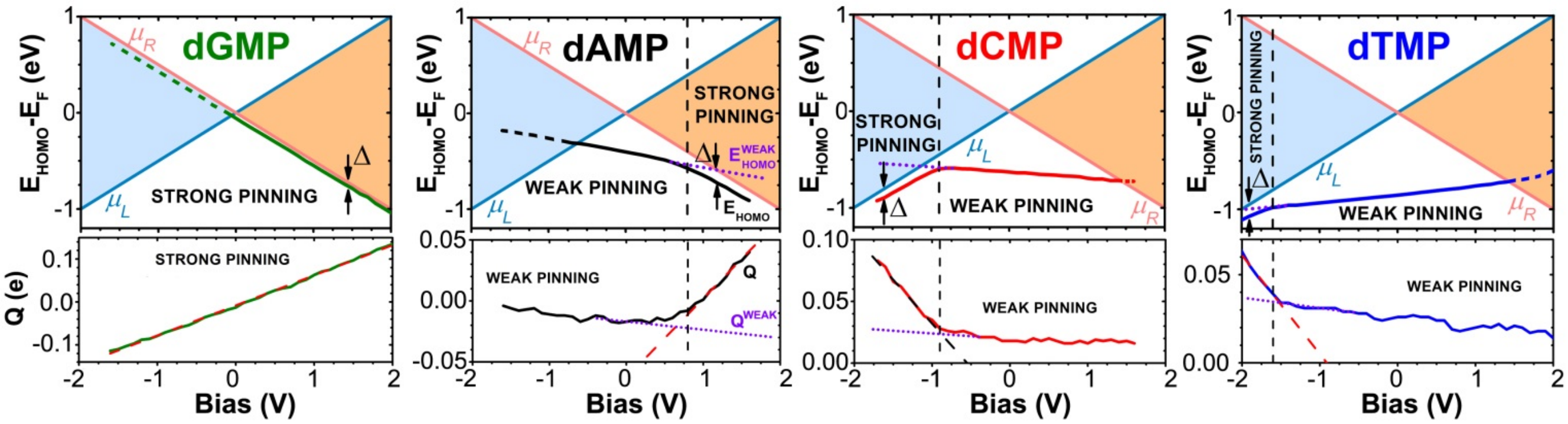

**Figure S7. Bias-dependent HOMO energy and Hirshfeld charge excess of nucleotides.** Top: The HOMO energy with respect to bias. The bias window for positive (negative) bias is shaded orange (light blue), while the chemical potential energies $\mu_R$ ($\mu_L$) are represented with orange (blue) solid lines. Dashed lines represent regions in which HOMO carries the current (enters the bias window). Vertical dashed lines for dAMP, dCMP and dTMP mark the boundary between weak and strong pinning regimes. Dotted purple $E_{HOMO}{}^{WEAK}$ lines are linear extrapolations of HOMO energy relative to $E_F$ in the weak pinning regime. The bias-independent distance $\Delta$ between HOMO peak energy and electrochemical potentials of electrodes is indicated with vertical arrows. Bottom: Charge excess $Q$ obtained from Hirshfeld analysis for each nucleotide versus bias. Linear fits (dashed red lines, black for dCMP) were performed on the strong pinning parts of the curves. Dotted purple lines $Q^{WEAK}$ are linear extrapolations of $Q$ from the weak pinning regime.

Shifting of transmission peaks with changing bias is best represented in the manner shown in the top panel of Fig. S7. Peak positions are depicted by solid coloured lines. Shaded areas represent bias windows, bordered by electrode electrochemical potential lines. The HOMO levels contribute to transport at bias values of around –0.7 V (dAMP) and around +1.5 V (dCMP, dTMP) and above (dashed lines). For opposite bias the HOMO peaks stay out of the bias window. We observe two distinct transmission regimes: strong and weak pinning. In the strong pinning regime, the HOMO peak retains a constant energy difference to the electrochemical potential of an electrode, shifting linearly with bias. For all nucleotides the HOMO peak remains outside of the bias window in the strong pinning regime, indicating non-resonant transport, except in the case of dGMP where by exception HOMO enters the negative bias window. dGMP is exceptional in the fact that it is always in the strong pinning regime. Other nucleotides exhibit both weak and strong pinning. Strong pinning of levels of dAMP to $\mu_R$ starts at +0.8 V, while those of dCMP and dTMP to $\mu_L$ starts at –0.9 V and –1.6 V, respectively. In the weak pinning regime, the transmission peak also shifts with bias, but to a lesser extent, and is not obviously pinned to the electrode electrochemical potential. This regime eventually leads to the HOMO peak entering the bias window, resulting in resonant transport. A total absence of pinning would result in the peak-positions curve parallel to the horizontal axis. To make clear the distinction between the two regimes we extrapolate the peak positions for the weak pinning regime, depicted by dotted purple lines in the top panel of Fig. S7.

The bottom panel of Fig. S7 depicts Hirshfeld charge excess $Q$ on a nucleotide with respect to bias. Finite charge excess at the nucleotide appears as result of charge redistribution between the molecule and electrodes, while the total considered system remains electroneutral. The difference between the two transport regimes is evident in the charge excess graph as well: charge excess is weakly dependent on bias for weak pinning, rising at a steep slope for strong pinning. This implies that strong pinning coincides with charge accumulation at the molecule.

## S5. Molecular charging

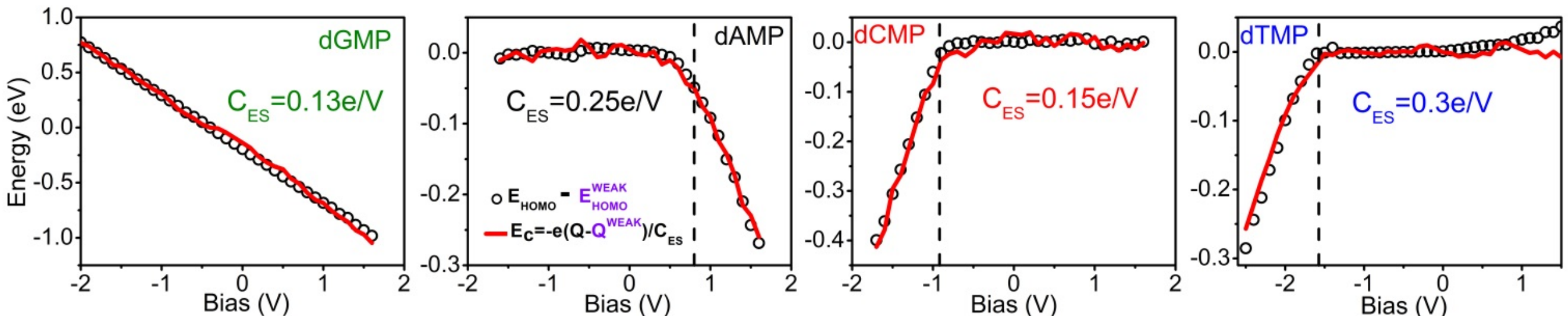

**Figure S8. Charging energy *$E_C$* (open circles) of nucleotides with respect to bias,** calculated from the difference between Hirshfeld charge excess $Q$ and its linear extrapolation $Q^{WEAK}$ (for dGMP $Q^{WEAK}$ is equal to $Q$ at zero bias) in the hypothetical case of weak pinning only, taking $C_{ES}$ to be equal to 0.13, 0.25, 0.15 and 0.3 e/V for dGMP, dAMP, dCMP and dTMP, respectively. The solid red line is the difference between the nucleotide $E_{HOMO}$ and its linear extrapolation $E_{HOMO}{}^{WEAK}$ in the hypothetical case of weak pinning only. For dGMP $E_{HOMO}{}^{WEAK}$ is equal to $E_{HOMO}$ at zero bias. The vertical dashed line marks the boundary between the weak and strong pinning regimes.

To quantify the effect that strong pinning has on the molecule, we find the charging energy $E_C$ as the difference between the energy that the HOMO peak would be positioned at in the hypothetical case of weak pinning only, $E_{HOMO}{}^{WEAK}$ (dotted purple lines in the top panel of Fig. S7), and actual $E_{HOMO}$ (Fig. S8 solid red lines). In the case of dGMP, which exhibits strong pinning for all explored biases the value of $E_{HOMO}{}^{WEAK}$ is taken to be equal to $E_{HOMO}$ at zero bias. The charging energy $E_C$ is also defined as:

$$E_C = -e(Q - Q^{WEAK})/C_{ES}$$

Where $e$ is elementary charge, $Q^{WEAK}$ is the linear extrapolation of $Q$ (for dGMP $Q^{WEAK}$ is equal to value of $Q$ at zero bias) in the hypothetical case of weak pinning only (dotted purple lines in the bottom panel of Fig. S7) and $C_{ES}$ is electrostatic capacitance across the system. $C_{ES}$ was used as a fitting parameter in the equation $E_{HOMO}{}^{WEAK}$ - $E_{HOMO}$ = $E_C$ (Fig. S8 open circles). We obtain a good fit for $C_{ES}$ values of 0.13, 0.25, 0.15 and 0.3 e/V for dGMP, dAMP, dCMP and dTMP, respectively, indicating that molecular charging is indeed the cause of the transition to strong pinning.

## S6. In-gap electrostatic potential and molecular HOMO wavefunctions

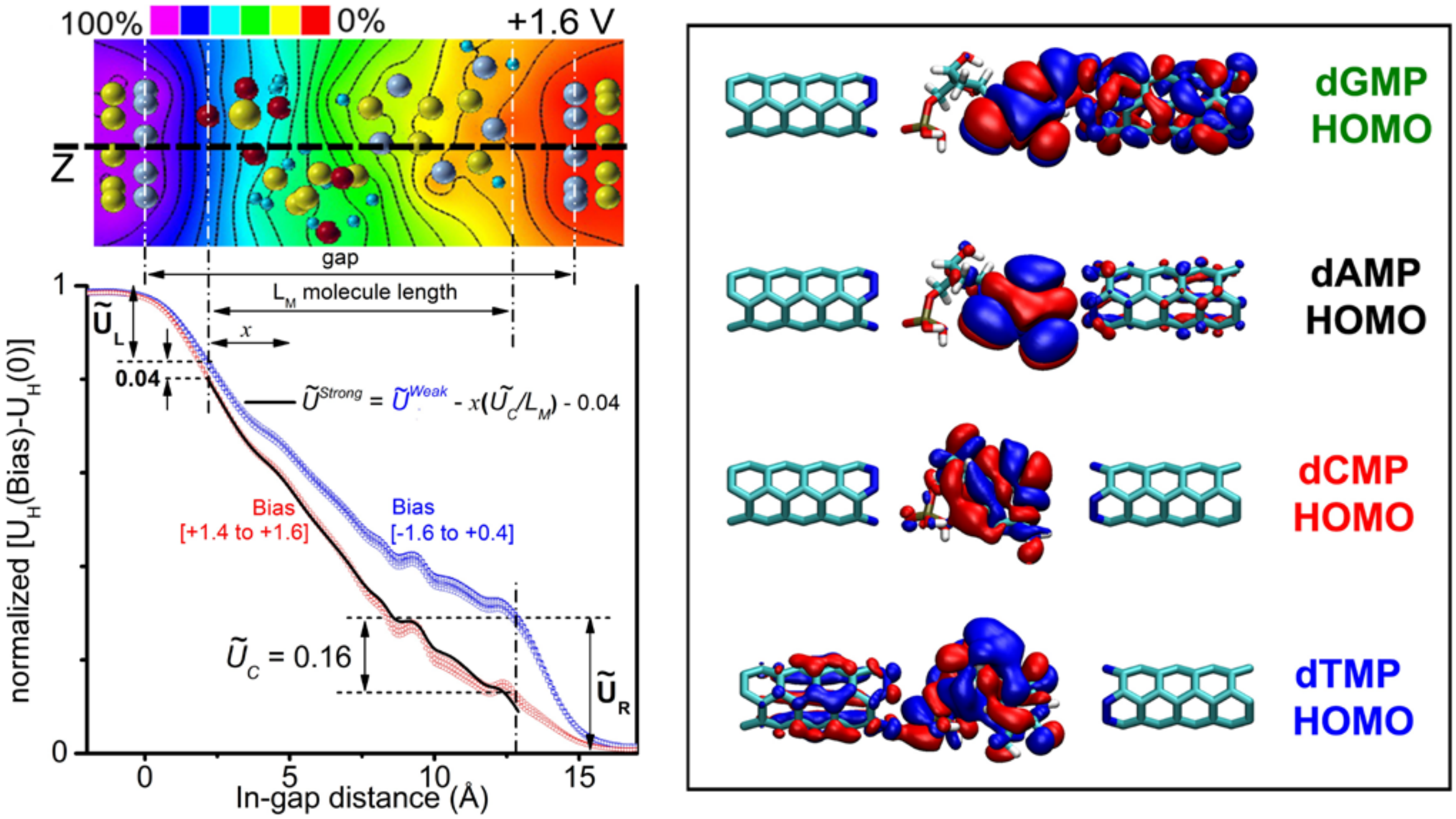

**Figure S9. In-gap electrostatic potential and molecular HOMO wavefunctions.** Left panel shows normalized electrostatic potential $U$(Bias)-$U$(0) extracted along the $z$-direction (black dashed line in top panel). Blue circles - bias range [-1.6, 0.4] and red circles - [+1.4, +1.6]. For clarity blue curves at negative bias are -($U$(Bias)-$U$(0)). $\tilde{U}_L$ and $\tilde{U}_R$ are normalized potential drops at left and right interfaces. $L_M$ is the length of molecule. The difference $\tilde{U}_C$ between mean values of blue and red curves is found to be equal to 0.16. The solid black line $\tilde{U}^{Strong} = \tilde{U}^{Weak} - x(\tilde{U}_C/L_M)$ - 0.04 is the linear transformation of the mean value $\tilde{U}^{Weak}$ of blue curves. Right panel shows spatial distribution of the wavefunction of the dGMP, dAMP, dCMP and dTMP HOMO levels at zero bias.

Spatial distribution of the nucleotide HOMO wavefunction at zero bias is given in Fig. S10. The HOMO is mainly localized at the nucleobase, except for dTMP. Spatial proximity leads to a spreading of the nucleotide's HOMO across to an electrode (to the right electrode for dGMP and dAMP, to the left electrode for dTMP). Such distribution does not significantly change with bias, and determines whether HOMO will be strongly pinned to the electrochemical potential of the left or right electrode. From Fig. S9 it is not obvious that dCMP HOMO will be pinned to the electrochemical potential of the left electrode. However, while there is no spreading of HOMO to the left electrode, HOMO is spatially closer to the left than to the right.

## S7. Influence of molecule rotation on transmission

Influence of nucleotide rotation (dGMP is given as an example) on transmission is shown in Fig. S11. When dGMP is rotated by 30° peak transmission diminishes by nearly two orders of magnitude which directly reflects on the current. However the position of HOMO with respect to the Fermi energy remains the same, implying that the rectifying ratio will also be insensitive to molecular rotation.

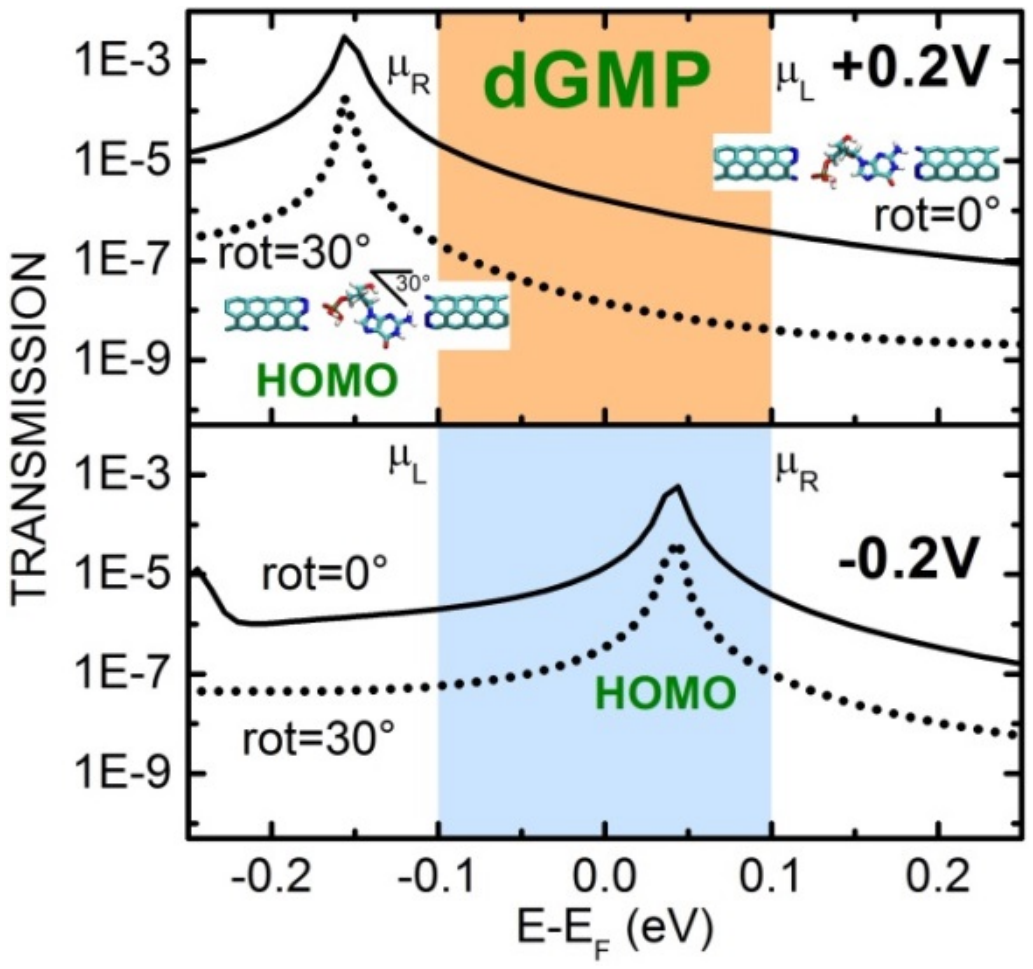

**Figure S10. Electronic transmission of dGMP rotated by 30° around the Y axis** (axis perpendicular to the plane that contains the nucleotide base, see inset). Top panel: transmission on logarithmic scale for dGMP (solid black line) and dGMP rotated by30° (dotted black line) at +0.2V. Bottom panel: transmission on logarithmic scale for dGMP (solid black line) and dGMP rotated by 30° (dotted black line) at -0.2V.